\documentclass{ws-p8-50x6-00}

\begin{document}

\newcommand{\beq}{\begin{equation}}
\newcommand{\eeq}{\end{equation}}
\newcommand{\beaq}{\begin{eqnarray}}
\newcommand{\eeaq}{\end{eqnarray}}
\newcommand{\la}{\langle }
\title{Chapman-Enskog expansion of the Boltzmann equation and
its diagrammatic interpretation}
\author{ M.E. Carrington${}^{a,b}$, Hou Defu${}^{a,b,c}$ and R.
Kobes${}^{b,d}$}

 \address{ ${}^a$ Department of Physics, Brandon University, Brandon, 
MB,R7A 6A9 Canada\\
 ${}^b$  Winnipeg Institute for Theoretical Physics, Winnipeg, Canada \\
${}^c$ Institute of Particle Physics, Huazhong Normal University, 430070 Wuhan,
China \\
${}^d$ University of Winnipeg, Winnipeg, Manitoba, R3B 2E9 Canada }

\maketitle

\abstracts{
We perform a Chapman-Enskog expansion of the Boltzmann equation 
keeping up to quadratic contributions.  We obtain a generalized nonlinear 
Kubo formula, and a set of integral equations which resum ladder and extended ladder diagrams. We show that these two equations have exactly the same structure, and thus provide a diagrammatic interpretation of the Chapman-Enskog expansion of the Boltzmann equation, up to quadratic order.}



    Fluctuations occur in a system perturbed slightly away from equilibrium.
The responses to these fluctuations are described by transport coefficients
 \cite{Groot}.
The investigation of transport coefficients in high temperature
gauge theories is important in cosmological applications
such as electroweak baryogenesis and in the context of heavy ion
collisions .
 There are two basic methods to calculate transport coefficients: transport
theory and linear response theory
\cite{MartinK,kubo,Hosoya}. 
To date, most calculations of transport coefficients have 
been done to the order of linear 
response. In many physical situations however nonlinear response can be 
important. In this talk, we study nonlinear response using transport theory quantum field theory, and explain the connection between these approaches.

 In a system that is out of 
equilibrium, the existence of gradients
in thermodynamic parameters give rise to thermodynamic forces which
lead to  deviations from the equilibrium expectation value of the
viscous shear stress:
\bea
&& \delta \langle  \pi_{\mu\nu}\rangle =\eta^{(1)} H_{\mu\nu} + \eta^{(2)}
H^{T2}_{\mu\nu} + \cdots  \label{DEF} \\
&& H_{\mu\nu} = \partial_\mu u_\nu + \partial_\nu u_\nu - \frac{2}{3}
\Delta_{\mu\nu} \Delta_{\rho\sigma}
\partial^\rho  u^\sigma,  \,\,\,
H_{\mu\nu}^{T2}:=H_{\mu\rho}H^\rho_{~\nu}-
\frac{1}{3}\Delta_{\mu\nu}H_{\rho\sigma}
H^{\rho\sigma} \nonumber
\eea
where
$u_\mu(x)$ is the four dimensional
four-velocity field which satisfies $u^\mu(x) u_\mu(x)=1$.
 The first
coefficient  is the usual shear viscosity.  The second has has not been 
widely discussed in the literature -- we will call it the quadratic shear 
viscous response.

The  Boltzmann equation can be used to calculate transport
properties for weak coupling $\lambda \phi^4$ theory with zero chemical potential\cite{jeon1}.
 We introduce a phase space distribution function
$f(x,\underline{k})$ (the underlined momenta  are on
shell).  The form of
$f(x,\underline{k})$ in local equilibrium is,
\beq
f^{(0)}=\frac{1}{e^{\beta(x) u_\mu(x) 
\underline{k}^\mu}-1}:=n_k\,;~~N_k:=1+2n_k
\,.\label{fequib}
\eeq
We expand $ f$  around $f^0$ using a
gradient expansion in the local rest frame where
$\vec{u}(x)=0$.   We keep only linear terms that contain one
power of $H_{\mu\nu}$ and quadratic terms that contain two powers of 
$H_{\mu\nu}$,. We write,
$f=f^{(0)} + f^{(1)} + f^{(2)} + \cdots$.
The viscous shear stress  tensor is given by
\beaq
\langle  \pi_{ij}\rangle  = \int \frac{d^3 k}{(2\pi)^3 2\omega_k}  f\, (k_i
k_j-\frac{1}{3}\delta_{ij} k^2)\,.
\eeaq
Using the gradient expansion of $f$ to calculate $\langle  \pi_{ij}\rangle$
and comparing with (\ref{DEF}) we have, 
\beaq
&&\eta^{(1)} = \frac{\beta}{15} \int \frac{d^3 k}{(2\pi)^3 2\omega_k}
n_k(1+n_k) k^2B(\underline{k}) \label{ttf11} \\
&& \eta^{(2)} = \frac{2\beta^2}{105}  \int \frac{d^3 k}{(2\pi)^3 2\omega_k}
[n_k(1+n_k)N_k] k^2  C(\underline{k})\,.\label{ttf}
\eeaq

    We show that $B(\underline{k})$ and
$C(\underline{k})$ can be
obtained from the first two equations in the hierarchy of equations obtained
from the Enskog expansion of the Boltzmann equation.
The first order equation can be cast into \cite{jeon1,meg},
\beaq
I_{ij}(k)=\frac{1}{2} \int_{123} d \,\Gamma_{12\leftrightarrow 3k}
d_n [B_{ij}(\underline{p}_1) 
+ B_{ij}(\underline{p}_2) - B_{ij}(\underline{k}) - B_{ij}(\underline{p}_3)\,]
\label{Bint}
\eeaq
where $d_n=(1+n_1)(1+n_2)n_3/(1+ n_k)$.
The second order Bolzmann equation leads to\cite{meg},
\beaq
  && N_kI_{ij}(k) B_{lm}(\underline{k}) = \frac{1}{2} \int_{123} d
\,\Gamma d_n\,\{ [ N_1 C_{ijlm}(\underline{p}_1)+N_2 C_{ijlm}(\underline{p}_2) - N_k C_{ijlm} (\underline{k})
\nonumber
\\
&& - N_3
C_{ijlm}(\underline{p}_3) ]
+\frac{1}{2}
[ N_{12}B_{ij}(\underline{p}_1)B_{lm}(\underline{p}_2)
-N_{k3}B_{ij}(\underline{p}_3)B_{lm}(\underline{k})
+\tilde N_{31}B_{ij}(\underline{p}_1)B_{lm}(\underline{p}_3)
\nonumber
\\
&&
+\tilde N_{k1}B_{ij}(\underline{p}_1)B_{lm}(\underline{k})+
\tilde N_{32}B_{ij}(\underline{p}_3) B_{lm}(\underline{p}_2)
+\tilde N_{k2}B_{ij}(\underline{k})B_{lm}(\underline{p}_2) ] \}\nonumber
\label{Cint}
\eeaq
where we used $N_{ij}=N_i+N_j$, $\tilde N_{ij}=N_i-N_j$ $(i,j=1,2,3,k)$. 
This  equation can be solved self consistently for the quantity
$C_{ijlm}(\underline{k})$ using the result for $B_{ij}(\underline{k})$ from
(\ref{Bint}).

Now we turn to response calculation \cite{Hosoya}. We work with the
density matrix in the Heisenberg representation
which satisfies
$\frac{\partial \rho}{\partial t}=0$ and  write $\rho=e^{-A+B}/ {\rm Tr} e^{-A+B}$
where $A =\int d^3 x F^\nu T_{0\nu}$ and
$B(t)=\int d^3 x \int _{-\infty}^t dt' e^{\epsilon (t'-t)}T_{\mu\nu}(x,t')
\partial^\mu F^\nu(x,t')$
with $F^\mu=\beta u^\mu$ and $\epsilon$ to be taken to zero at the end. Here
 $A$ is the equilibrium part of the Hamiltonian and $B$ is a
perturbative contribution. We expand  the density matrix in $B$ and find
 the shear viscosity 
\beaq
\eta^{(1)} = \frac{1}{10}\frac{d}{d q_0}{\rm Im} [ \lim_{\vec{q} \to
0}D_R(Q)]|_{q_0=0}\,. \label{pl}
\eeaq
This is the well known Kubo formula \cite{kubo,Hosoya}.
  The  quadratic shear viscous
result can be written as a
retarded three-point correlator \cite{mhrkb}:
\beaq
\eta^{(2)} = \frac{3}{70}\frac{d}{d q_0} \frac{d}{ dq_0'}{\rm Re}\,[
\lim_{\vec{q} \to 0}G_{R1}(-Q-Q',Q,Q')]|_{q_0=q'_0=0} \nonumber
\label{pq}
\eeaq
  We have obtained a type of nonlinear Kubo
formula that allows us to obtain the quadratic shear viscous response from a
retarded three-point function using equilibrium quantum field theory.

Next
we obtain a perturbative expansion for the correlation functions of
composite operators
 $D_R(x,y)$ and $G_{R1}(x,y,z)$ which appear in (\ref{pl}) and (\ref{pq}).
We use the CTP formulation  and work in the Keldysh representation.  We
define the vertices
$\Gamma_{ij}$ and $M_{ijlm}$ by
truncating external legs from the following connected vertices:
$\Gamma^C_{ij} = \langle T_c \pi_{ij}(x) \phi(y) \phi(z)\rangle  
$, $ M^C_{ijlm} = \langle T_c \pi_{ij}(x) \pi_{lm}(y)\phi(z) \phi(w)\rangle
$
where $\pi_{ij}(x) = \partial_i\phi(x) \partial_j\phi(x) - \frac{1}{3} \delta_{ij}(\partial_m \phi(x))(\partial_m \phi(x))$.
These definitions allow us to write the two- and three-point correlation
functions as integrals of those vertices. Using the Kubo formulea above we
obtain\cite{meg},

\beaq
&& \eta^{(1)} = \frac{\beta}{15} \int dK\, k^2 \,\rho_k n_k (1+n_k) \left[\frac
{\rm Re \Gamma_{R2}(K)}{\rm Im \Sigma_k}\right] \label{ftf11} \\
&& \eta^{(2)} =  -\frac{2\beta^2}{105} \int dK \,k^2\rho_k n_k (1+n_k) N_k
\left[ \frac{\rm Re M_{R1}(K)}{\rm Im \Sigma_k} \right]\,. \label{ftf}
\eeaq
Comparing with (\ref{ttf11}) and (\ref{ttf}) we see that the results are
identical if we identify
\beaq
B(\underline{k}) = \frac {\rm Re \Gamma_{R2}
(\underline{k})}{\rm Im \Sigma_k},\, ~~~
 C(\underline{k}) = -  \frac{\rm Re M_{R1}(\underline{k})}{\rm Im \Sigma_k}
\label{EQ2}
\eeaq
with the momentum $K$ on the shifted mass shell: $\delta(K^2-m_{th}^2)$ where
$m_{th}^2=m^2+{\rm Re} \Sigma_K$.

It is well known that ladder diagrams  give the
dominate contributions to the vertex $\Gamma_{ij}$. They
 contribute to the viscosity to the same order in perturbation
theory as the
bare one loop graph and thus need to be included in a resummation.
The integral equation that one obtains from resumming ladder contributions 
to the three-point vertex 
has exactly the same form as the equation obtained from the
linearized Boltzmann equation (\ref{Bint}) with a shifted mass shell 
describing effective thermal excitations \cite{jeon1,meg}.

Following the pinch effect argument\cite{jeon1,mhrkb}, one can show that an
infinite set of ladder graphs and some other 
contributions which we will call extended ladder graph contribute
to the same order to  vertex   $M_{ijlm}$ as the tree diagram. Therefore,
for consistent calculation, we consider an integral equation which resums
all of these diagrams,  as shown in Fig. ~1.
\begin{eqnarray}
\parbox{14cm}
{{
\begin{center}
\parbox{10cm}
{
\epsfxsize=10cm
\epsfysize=3cm
\epsfbox{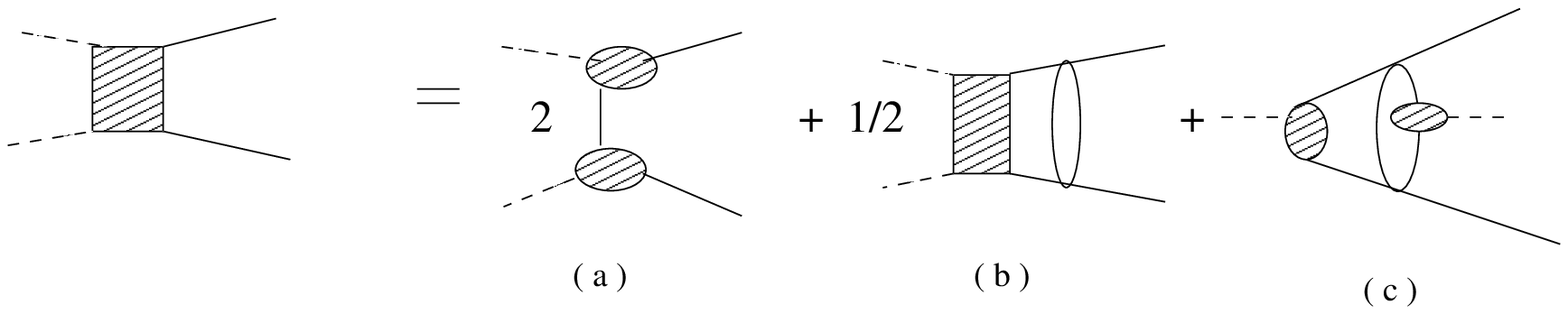}}\\
\parbox{10cm}{\small  Fig.~1:  Integral equation for an
extended-ladder resummation.  }
\label{F1}
\end{center}
}}
\nonumber
\end{eqnarray}

This Integral equation can be cast into\cite{mhrkb}:

\beaq
&& N_kI_{ij} \frac{\Gamma_{R2}^{lm}(K)}{{\rm Im}\Sigma_k} 
=\frac{\lambda^2}{12} \int_{123}{(2\pi)}^4\delta^4_{123K}
d_n \rho_1 \rho_2 \rho_3
[\frac{N_{p_3} M_{R1}^{ijlm} (P_3) }{{\rm Im}\Sigma_{p_3}}
\label{ieq2}
\\
&&
+\frac{N_{k} M_{R1}^{ijlm} (K) }{{\rm Im}\Sigma_k}
 - \frac{N_{p_1} M_{R1}^{ijlm} (P_1) }{{\rm Im}\Sigma_{p_1}}
- \frac{N_{p_2} M_{R1}^{ijlm} (P_2) }{{\rm Im}\Sigma_{p_2}}
\nonumber
\\
&&
+\frac{1}{2}\{ N_{12}\frac{\Gamma_{R2}^{ij}(P_1)}{{\rm
Im}\Sigma_{p_1}}
\frac{\Gamma_{R2}^{lm}(P_2)}{{\rm Im}\Sigma_{p_2}}
-N_{k3}\frac{\Gamma_{R2}^{ij}(K)}{{\rm Im}\Sigma_k}
\frac{\Gamma_{R2}^{lm}(P_3)}{{\rm Im}\Sigma_{p_3}}
+\tilde N_{31}\frac{\Gamma_{R2}^{ij}(P_1)}{{\rm Im}\Sigma_{p_1}}
\frac{\Gamma_{R2}^{lm}(P_3)}{{\rm Im}\Sigma_{p_3}}
\nonumber
\\
&&
+ \tilde N_{k1}\frac{\Gamma_{R2}^{ij}(P_1)}{{\rm Im}\Sigma_{p_1}}
\frac{\Gamma_{R2}^{lm}(K)}{{\rm Im}\Sigma_k}
+\tilde N_{32}\frac{\Gamma_{R2}^{ij}(P_3)}{{\rm Im}\Sigma_{p_3}}
\frac{\Gamma_{R2}^{lm}(P_2)}{{\rm Im}\Sigma_{p_2}}
+ \tilde N_{k2}\frac{\Gamma_{R2}^{ij}(K)}{{\rm Im}\Sigma_k}
\frac{\Gamma_{R2}^{lm}(P_2)}{{\rm Im}\Sigma_{p_2}}
    \}]\,
\nonumber
\eeaq
Note that once again we have obtained an integral equation that is decoupled:
it only involves $M_{R1}$ and $\Gamma_{R2}$.  With $\Gamma_{R2}$
determined from the integral equation for the ladder resummation, (\ref{ieq2})
can be solved to obtain $M_{R1}$.
  Finally, comparing (\ref{ftf}) and (\ref{ieq2}) with (\ref{ttf}) and
(\ref{Cint}) we see that calculating the quadratic shear viscous response using
 transport theory describing effective thermal excitations and keeping terms
that are quadratic in the gradient of the four-velocity field in the expansion
of the Boltzmann equation is equivalent to calculating the same response
coefficient from quantum field theory at finite temperature using the
next-to-linear response Kubo formula with a vertex given by a specific integral
equation. This integral equation shows that the complete set of diagrams that
need to be resummed includes the standard ladder graphs, and an additional set
of extended ladder graphs. Some of the diagrams that contribute to the
viscosity are shown in Fig.~2.

\begin{eqnarray}
\parbox{14cm}
{{
\begin{center}
\parbox{10cm}
{
\epsfxsize=10cm
\epsfysize=3cm
\epsfbox{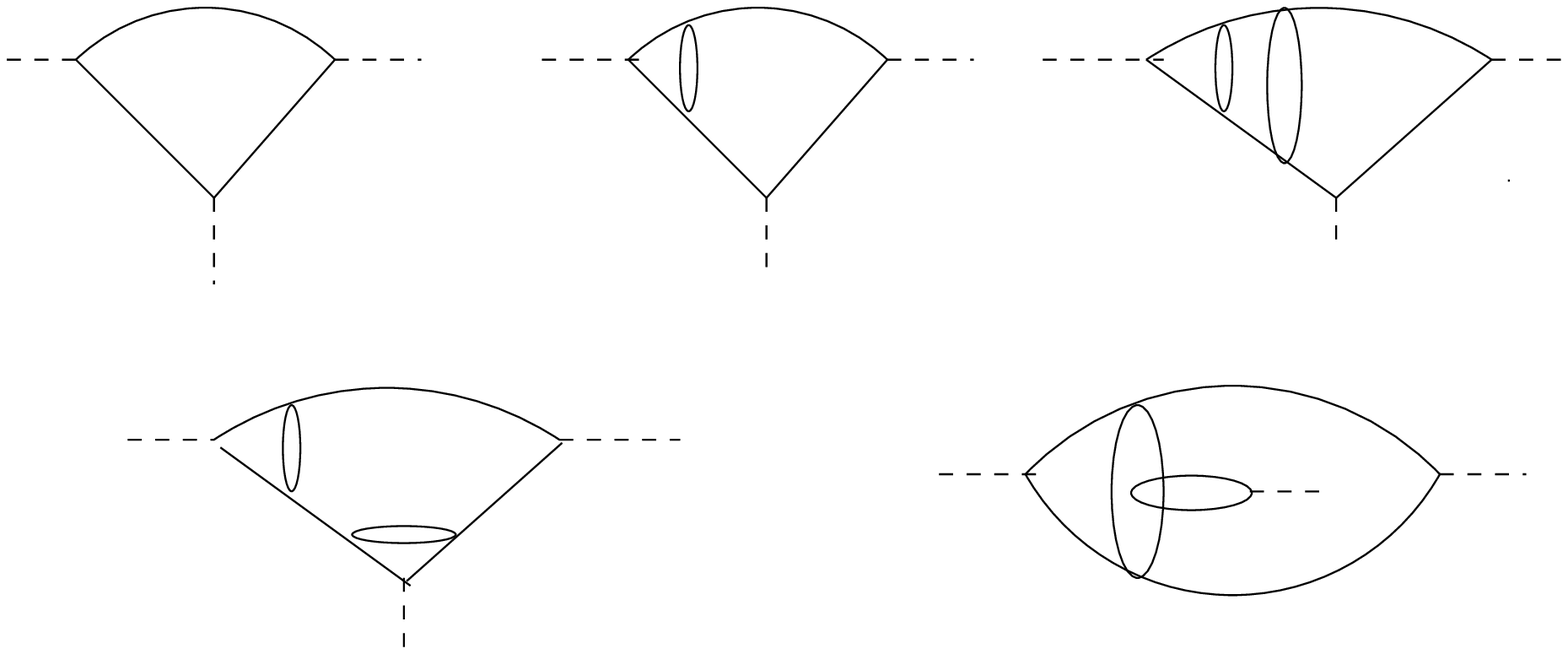}}\\
\parbox{10cm}{\small  Fig.~2:  Some of the ladder and extended ladder
diagrams that contribute to quadratic shear viscous response. }
\label{F7}
\end{center}
}}
\nonumber
\end{eqnarray}

This result provides a diagrammatic interpretation of the Chapman-Enskog 
expansion of Boltzmann equation, up to quadratic order.

\section*{Acknowledgments}
 This work was partly supported by NSFC and NSERC

\end{document}